\pdfoutput=1
\documentclass[11pt,aps,a4paper,
  amsmath,amssymb,
  notitlepage,nofootinbib,%endfloats*,
  longbibliography
  ]{revtex4-1}

\usepackage{color}

\usepackage{graphicx}

\begin{document}
\preprint{} 

\title{Spectral flow for an integrable staggered superspin chain}

\author{Holger Frahm}
\author{Konstantin Hobu{\ss}}
\affiliation{Institut f{\"u}r Theoretische Physik,
  Leibniz Universit{\"a}t Hannover  \\ 
  Appelstra{\ss}e 2, 30167 Hannover, Germany}

%\date{\today}

\begin{abstract}
  The flow of the low energy eigenstates of a $U_q[sl(2|1)]$ superspin chain
  with alternating fundamental ($3$) and dual ($\bar{3}$) representations is
  studied as function of a twist angle determining the boundary conditions.
  The finite size spectrum is characterized in terms of scaling dimensions and
  quasi momenta representing the two families of commuting transfer matrices
  for the model which are even and odd under the interchange $3\leftrightarrow
  \bar{3}$, respectively.  
  Based on the extrapolation of our finite size data we find that under a
  variation of the boundary conditions from periodic to antiperiodic for the
  fermionic degrees of freedom levels from the continuous part of the spectrum
  flow into discrete levels and vice versa.
\end{abstract}

\maketitle
%\section{Introduction}
%
Integrable two-dimensional vertex models and the corresponding
$(1+1)$-dimensional quantum spin chains have provided many examples for
lattice realizations of conformal field theories (CFTs) and thereby led to
important insights into the critical properties of strongly correlated
systems.
For lattice models with finite dimensional representations for the local
degrees of freedom and Hermitian Hamiltonian the unitary CFTs appearing in the
continuum limit can be identified by relating Virasoro representations to the
discrete finite size spectrum of the former \cite{BlCN86,Affl86}.

Recently, a growing number of lattice models related to order-disorder
transitions in quantum Hall systems, the anti-ferromagnetic Potts model,
intersecting loops or physical properties of two-dimensional polymers, has
been found whose continuum limit -- although having a finite-dimensional
quantum space for the local degrees of freedom on the lattice -- are CFTs with
a non-compact target space
\cite{EsFS05,IkJS08,FrMa11,FrMa12,VeJS14,FrMa15,VeJS16a}.  This manifests
itself in the emergence of a continuous component in the spectrum of scaling
dimensions in the thermodynamic limit and, possibly, a non-analytic dependence
of the effective central charge of the lattice model on the twist of boundary
conditions related to discrete levels appearing in the spectrum of the lattice
model when the corresponding operator in the CFT becomes normalizable.

For some of these lattice models the spectral data obtained from their exact
solution have allowed to identify the CFT describing their continuum limits:
\begin{itemize}
\item An integrable vertex model built from the three-dimensional fundamental
  and dual representations of the superalgebra $sl(2|1)$ was shown to flow to
  a $SU(2|1)$ Wess-Zumino-Novikov-Witten (WZNW) model at level $k=1$
  \cite{EsFS05,SaSc07}.  For a physical modular invariant partition function
  one is forced to consider continuous values of the $sl(2|1)$ charge quantum
  number leading to continua of scaling dimensions.
\item A staggered six-vertex model related to the anti-ferromagnetic Potts
  model \cite{IkJS08} (which also appears in the phase diagram of a staggered
  superspin chain built from four-dimensional representations of
  $U_q[sl(2|1)]$, see Ref.~\cite{FrMa12}) has been proven to provide a
  realization of the $SL(2,\mathbb{R})/U(1)$ Euclidean black hole sigma model
  \cite{Witten91,DiVV92}.  Here the identification was based on the density of
  states in the continuous spectrum \cite{IkJS12,CaIk13,FrSe14}.
\item The $a_{2}^{(2)}$ model (equivalent to the 19-vertex Izergin-Korepin model
  \cite{IzKo81}) in 'regime III' has a non-compact continuum limit again
  described by the $SL(2,\mathbb{R})/U(1)$ black hole sigma model
  \cite{VeJS14}.  Here the spectrum of discrete levels appearing in the
  lattice model with twisted boundary conditions has been found to be
  consistent with the predicted appearance of levels related to the principal
  discrete representations of $SL(2,\mathbb{R})$ states \cite{HaPT02,RiSc04}.
  Similarly, the scaling limit of the general $a_{N-1}^{(2)}$ model has been
  shown to be a $SO(N)/SO(N-1)$ gauged WZNW model \cite{VeJS16a}.
\end{itemize}

In this paper we are studying the continuum limit of the mixed $U_q[sl(2|1)]$
superspin chain based on the three dimensional atypical representation and its
dual (``quark'' and ``antiquark'' , labeled $3$ and $\bar{3}$ in
the following) \cite{Gade99}, an anisotropic deformation of the $sl(2|1)$
superspin chain mentioned above.  In previous work on this model, and similar
as in the isotropic case, the existence of an exact zero energy state and of
continua of scaling dimensions have been established \cite{FrMa11}.  Here we
generalize the lattice model to include
% twisted
a twist allowing for an adiabatic change from periodic to anti-periodic
boundary conditions for the fermionic degrees of freedom.  Using the Bethe
ansatz solution of this model we find the exact dependence of some low lying
levels on the twist angle.  Translating our results into the context of the
field theory describing the continuum limit of the model we find that under
the spectral flow states from the continuous part of the spectrum in the
Neveu-Schwarz sector are mapped onto discrete levels in the Ramond sectors,
and vice versa.

%\section{The mixed $U_q[sl(2|1)]$ superspin chain}
%\label{sec:superspinchain}
%
The construction of an integrable mixed superspin chain with
alternating three-dimensional quark and anti-quark representations of
$U_q[sl(2|1)]$ is based on four $\mathcal{R}$-matrices acting on
tensor products $3\otimes3$, $3\otimes\bar{3}$, $\bar{3}\otimes3$, and
$\bar{3}\otimes\bar{3}$ satisfying Yang-Baxter equations
\begin{equation}
  \label{ybe}
  \mathcal{R}_{12}^{(\omega_1,\omega_2)}(\lambda)
  \mathcal{R}_{13}^{(\omega_1,\omega_3)}(\lambda+\mu)
  \mathcal{R}_{23}^{(\omega_2,\omega_3)}(\mu) 
  = \mathcal{R}_{23}^{(\omega_2,\omega_3)}(\mu)
  \mathcal{R}_{13}^{(\omega_1,\omega_3)}(\lambda+\mu)
  \mathcal{R}_{12}^{(\omega_1,\omega_2)}(\mu) \,,
\end{equation}
with $\omega_j\in\{3,\bar{3}\}$ for $j=1,2,3$ \cite{Gade99} (see also
\cite{PeSc81,KuSk82a}).  Two families of row-to-row transfer matrices acting
on a Hilbert space $(3\otimes\bar{3})^{\otimes L}$ are constructed as the
supertrace over auxiliary spaces $A\cong\mathbb{C}^3$ of ordered products of
these $\mathcal{R}$-matrices as \cite{Kulish85}
\begin{equation}
  \label{eq:transfer}
  \begin{aligned}
  \tau^{(3)}(\lambda) &= \mathrm{str}_A\left( \mathcal{G}^{(3)}(\alpha)
    \mathcal{R}^{(3,3)}_{A,2L}(\lambda)
    \mathcal{R}^{(3,\bar{3})}_{A,2L-1}(\lambda - i\gamma)
    \mathcal{R}^{(3,3)}_{A,2L-2}(\lambda)
    \ldots \mathcal{R}^{(3,\bar{3})}_{A,1}(\lambda - i\gamma)\right)\,,
  \\
  \tau^{(\bar{3})}(\lambda) &= \mathrm{str}_A\left( \mathcal{G}^{(\bar{3})}(\alpha)
    \mathcal{R}^{(\bar{3},3)}_{A,2L}(\lambda + i\gamma)
    \mathcal{R}^{(\bar{3},\bar{3})}_{A,2L-1}(\lambda)
    \mathcal{R}^{(\bar{3},3)}_{A,2L-2}(\lambda + i\gamma)
    \ldots \mathcal{R}^{(\bar{3},\bar{3})}_{A,1}(\lambda)\right)\,.
  \end{aligned}
\end{equation}
Here $\gamma$ parametrizes the deformation parameter
$q=\exp(-i\gamma)$, more details can be found in
Ref.~\cite{FrMa11}.\footnote{%
  Note, the definition of $\gamma$ in Ref.~\cite{FrMa11} differs from
  the one used here by a factor of 2.}
In Eq.~(\ref{eq:transfer}) boundary conditions are controlled by the diagonal
twist matrices $\mathcal{G}^{(\omega)}(\alpha) = \exp\left(2i\alpha
  J_3^{(\omega)}\right)$, $J_3^{(\omega)}$ being the diagonal generator of the
spin-subalgebra in the representation $\omega$.  For $\alpha=0$ the lattice
model obeys periodic boundary conditions while for $\alpha=\pm\pi$ bosons are
periodic while fermions obey \emph{anti}periodic boundary conditions.  In the
field theory describing the continuum limit of the lattice model these cases
correspond to the Ramond (R) and Neveu-Schwarz (NS) sector, respectively.
 
As a consequence of (\ref{ybe}) the two transfer matrices commute,
$\left[ \tau^{(\omega_1)}(\lambda), \tau^{(\omega_2)}(\mu) \right]=0$ for
all $ \lambda$, $\mu$ and $\omega_j\in\{3,\bar{3}\}$.
Local integrals of motion are generated by the double row transfer matrix
\begin{equation}
  \label{eq:drtransfer}
  \tau(\lambda) = \tau^{(3)}(\lambda) \tau^{(\bar{3})}(\lambda)\,.
\end{equation}
For example, the Hamiltonian of the mixed $U_q[sl(2|1)$ superspin chain is
\begin{equation}
  \label{eq:hamiltonian}  
  \mathcal{H} = i\frac{\partial}{\partial \lambda} 
    \ln  \tau(\lambda) \Big|_{\lambda=0}\,.
\end{equation}
%----------------------------------------------------------------------

For a staggered model as the one considered here we can introduce another
combination of the single-row transfer matrices, i.e.\ $\tau^{(3)}(\lambda)
[\tau^{(\bar{3})}(\lambda)]^{-1}$, whose logarithm, unlike
(\ref{eq:drtransfer}), is \emph{odd} under the exchange of the quark and
antiquark representation.  Among the conserved quantities generated by this
object, and motivated by the analysis of a staggered six-vertex model by Ikhlef
\emph{et al}.  \cite{IkJS12}, see also Refs.~\cite{CaIk13,FrSe14}, we 
consider the `quasi momentum operator'
\begin{align}
  \label{eq:quasimom}
  \mathcal{K} &= 
  \frac{\gamma}{2\pi(\pi-2\gamma)}\,
  \ln \left( \tau^{(3)}(\lambda)
    \left[\tau^{(\bar{3})}(\lambda) \right]^{-1} \right)
  \Big|_{\lambda = 0}\,.
\end{align}

%----------------------------------------------------------------------
The transfer matrices (\ref{eq:transfer}) (and therefore $\mathcal{H}$
and $\mathcal{K}$) can be diagonalized using the nested algebraic Bethe
ansatz \cite{KuRe83, BaVV82}.  The resulting expressions obtained
within this framework depend on the choice of grading for the
underlying superalgebra \cite{Kulish85,EsKo92,FoKa93,PfFr96,PfFr97}.
As in Refs.~\cite{EsFS05,FrMa11} we use $[p_1,p_2,p_3] = [0,1,0]$.
In the sector of the Hilbert space with fixed quantum numbers related
to the $U(1)$ subalgebras of $U_q[sl(2|1)]$, i.e.\ charge $b = (N_1 -
N_2)/2$ and $z$-component of the spin $j_3 = L - (N_1+N_2)/2$ with
non-negative integers $N_{1,2}$ the spectrum of the transfer matrices
can be parametrized $N_1+N_2$ complex rapidities
$\{\lambda_j^{(1)}\}_{j=1}^{N_1}$ and
$\{\lambda_j^{(2)}\}_{j=1}^{N_2}$ solving the Bethe equations,
\begin{equation}
\label{eq:BAE}
\begin{split}
  \left[ \frac{\sinh (\lambda_j^{(1)} + i\gamma) }{\sinh
      (\lambda_j^{(1)} - i\gamma)}\right]^L &= \mathrm{e}^{i\alpha}
  \prod
  \limits_{k=1}^{N_2} \frac{\sinh (\lambda_j^{(1)} - \lambda_k^{(2)} +
    i\gamma)}{\sinh (\lambda_j^{(1)} - \lambda_k^{(2)} - i\gamma)}\,,
  \quad j=1,\ldots,N_1\,,
  \\
  \left[ \frac{\sinh (\lambda_j^{(2)} + i\gamma) }{\sinh
      (\lambda_j^{(2)} - i\gamma)}\right]^L &= \mathrm{e}^{i\alpha}
  \prod
  \limits_{k=1}^{N_1} \frac{\sinh (\lambda_j^{(2)} - \lambda_k^{(1)} +
    i\gamma)}{\sinh (\lambda_j^{(2)} - \lambda_k^{(1)} - i\gamma)}\,,
  \quad j=1,\ldots,N_2\,.
\end{split}
\end{equation}
The corresponding eigenvalues of the Hamiltonian (\ref{eq:hamiltonian})
and the quasi momentum (\ref{eq:quasimom}) are
\begin{equation}
  \label{eig-Bethe}
  \begin{aligned}
    E(\{\lambda_j^{(1)}\},\{\lambda_j^{(2)}\})% &
    % = i
    % \frac{\partial}{\partial \lambda}\ln \Lambda(\lambda) \Big|_{\lambda
    %   = 0}
    % \\
    &= 4L \cot(\gamma) + 2\left( \sum \limits_{k=1}^{N_1} \frac{\sin
        (2\gamma)}{\cos (2\gamma) - \cosh (2 \lambda_k^{(1)}) } + \sum
      \limits_{k=1}^{N_2} \frac{\sin (2\gamma)}{\cos (2\gamma) - \cosh
        (2 \lambda_k^{(2)}) } \right)\,,
    \\
    K(\{\lambda_j^{(1)}\},\{\lambda_j^{(2)}\}) 
    &= \frac{\gamma}{2\pi(\pi-2\gamma)}\,\left(
        \sum \limits_{k=1}^{N_1} \ln \left( \frac{\sinh(
        \lambda_k^{(1)} + i\gamma )}{\sinh( \lambda_k^{(1)} - i\gamma
        )} \right) - \sum \limits_{k=1}^{N_2} \ln \left( \frac{\sinh(
        \lambda_k^{(2)} + i\gamma )}{\sinh( \lambda_k^{(2)} - i\gamma
        )} \right)
  \right)\,.
  \end{aligned}
\end{equation}

%%%%%%%%%%%%%%%%%%%%%%%%%%%%%%%%%%%%%%%%%%%%%%%%%%%%%%%%%%%%%%%%%%%%%%% 
%\section{Finite size spectrum of the superspin chain}
%\label{sec:fsspectrum}
%
Previous studies of this model \cite{EsFS05,FrMa11,FrMa12} have shown that the
spectrum of the transfer matrix (\ref{eq:drtransfer}) in the sector with
charge $b=0$ (i.e.\ $N_1=N_2=N$) contains (after scaling by a factor of two)
that of the antiferromagnetic spin-$1$ XXZ Heisenberg model with twist
$\varphi=\pi+\alpha$ \cite{ZaFa80}: for a subset of eigenstates of the mixed
superspin chain the Bethe roots degenerate as
$\lambda_j^{(1)}=\lambda_j^{(2)}=\lambda_j$.  These corresponding eigenvalues
can be identified with eigenenergies of the spin-$1$ chain in the sector with
$j_3=L-N$, see Refs.~\cite{FrMa11,FrMa12}.\footnote{%
  The eigenvalues $K$ of the quasi momentum operator, Eq.~(\ref{eig-Bethe}),
  are odd under the interchange $\{\lambda_j^{(1)}\} \leftrightarrow
  \{\lambda_j^{(2)}\}$ changing the sign of the charge $b$.  As an immediate
  consequence $K = 0$ for the levels from the spin-$1$ XXZ chain subset of the
  spectrum.}
At low energies the operator content of the effective field theory for the
spin-$1$ XXZ model is given in terms of composites of a $U(1)$ Kac-Moody field
and Ising operators \cite{AlMa89,AlMa90,FrYF90}.  This provides a subset of
the finite size spectrum of the superspin chain \cite{EsFS05,FrMa11}.  
After rescaling of the finite size energy gaps, $\Delta E \to (L\Delta E/2\pi v_F)
 \equiv X$ ($v_F=\pi/\gamma$ is the Fermi velocity of the gapless
excitations), the corresponding effective scaling dimensions of the superspin
chain (\ref{eq:hamiltonian}) are
% The effective scaling dimensions of the superspin chain
% (\ref{eq:hamiltonian}) are ($v_F=\pi/\gamma$ is the Fermi velocity of the
% gapless excitations)
%
\begin{equation}
  \label{eq:fsspec}
  \begin{aligned}
  X^{\mathrm{eff}}_{(m,w)}(\varphi=\alpha+\pi) %&= \frac{L}{2\pi\,v_F} E_{(m,w)}\\
  &= -\frac14\,\delta_{m+w\in2\mathbb{Z}} 
  + \frac{m^2}{2k} + \frac{k}{2}
  \,\left(w+\frac{\varphi}{\pi}\right)^2\,,
  \qquad k=\frac{\pi}{\pi-2\gamma}\,,
\end{aligned}
\end{equation}
where the integer $m$ is the quantum number $j_3$ of the corresponding state
in the superspin chain while $w$ is related to the vorticity of that state.
Note that the energy of the level $(m,w)=(0,-1)$ in the Ramond sector
($\varphi=\pi$) vanishes identically.  The Bethe root configuration
corresponding to this exact zero mode in the $(b,j_3)=(0,0)$ sector of the
periodic superspin chain is highly degenerate, i.e.\ $\lambda_j^{(a)}\equiv 0$
for all $j=1,\ldots,L$ and $a=1,2$.  This is the ground state of the periodic
superspin chain (and the related spin-$1$ XXZ model with anti-periodic
boundary conditions where the zero-mode has been shown to be a singlet under
an exact dynamical lattice supersymmetry \cite{Hagendorf15}) for anisotropies
$0\le\gamma\le\pi/4$.

In the large $L$ limit the possible root configurations solving the Bethe
equations (\ref{eq:BAE}) can be classified using the string hypothesis.  A
class of low energy excitations in the zero charge sector $b=0$ has been
identified with collections of $O(L)$ 'strange $2$-strings' consisting of
complex conjugate pairs of rapidities
$\lambda^{(1)}=\left(\lambda^{(2)}\right)^*$.  These strange strings come in
two types, i.e.\ $(\pm)$ with $\mathrm{Im}(\lambda^{(1)}) = \pm \gamma/2$.
Solutions to the Bethe equations consisting of $N_\pm$ type-$(\pm)$ strange
strings with $N_++N_-=L-j_3$ but $\Delta N = N_+-N_- \ne 0$ have been found to
form a continuous component of the finite size spectrum starting at levels
% (\ref{eq:fsspecR}), (\ref{eq:fsspecNS}) 
(\ref{eq:fsspec}) with $m+w\in2\mathbb{Z}$ \cite{EsFS05,FrMa11}.

Based on these insights from the lattice model it has been argued that the
continuum limit of the isotropic (i.e.\ $\gamma=0$) superspin chain flows to a
$SU(2|1)$ WZNW model at level $k=1$ \cite{EsFS05,SaSc07}.  The generic class~I
irreducible representations of the affine superalgebra
$\widehat{sl}(2|1)_{k=1}$ are built over the typical representations
$[b,j=\frac12]$ with charge $b\in\mathbb{C}$ and spin $j=\frac12$.  To obtain
a physical modular invariant partition function from the characters for these
representations which can be compared with the spectrum of the lattice model
one is forced to consider continuous values of the charge $b$ which,
after analytical continuation $b\to i\beta$ does yield continua starting at
scaling dimensions $X^{\mathrm{eff},R}(m,w) \equiv
X^{\mathrm{eff}}_{(m,w)}(\varphi=\pi) \in \mathbb{N} + \frac14$ in the Ramond
sector \cite{SaSc07}.
Note, however, that this proposal does not capture the primary fields with
integer scaling dimension $X^{\mathrm{eff},R}$ (i.e.\ $m+w$ odd in
(\ref{eq:fsspec})), in particular the $sl(2|1)$ singlet state with
$X^{\mathrm{eff},R}=0$.  Since modular invariance appears to preclude the
appearance of this singlet on its own it has been argued in Ref.~\cite{SaSc07}
that this state is an artifact of the lattice model which disappears in the
continuum limit once the partition function is properly normalized.

%----------------------------------------------------------------------
Here we will not try to extend this picture to describe the continuous
spectrum of critical exponents of the anisotropic superspin chain.
Instead we focus on another characteristic feature of certain
conformal field theories with non-compact target space, i.e.\ the
appearance of a discrete component of the spectrum.  To be specific we
study the spectrum of the $U_q[sl(2|1)]$ superspin chain as a function
of the twist $\varphi=\alpha+\pi$ which, in the continuum limit,
corresponds to the spectral flow between the Neveu-Schwarz sector and
the Ramond sector.

%----------------------------------------------------------------------
The ground state of the former is found in the sector with quantum numbers
$(b,j_3)=(0,0)$.  For even $L$ it is realized in the XXZ subspace
of this sector and the corresponding Bethe root configuration can be mapped to
that of the singlet ($j_3=0$) ground state of the periodic spin-$1$ chain.
Its effective scaling dimension is $X^{\mathrm{eff},NS}(0,0)\equiv
X^{\mathrm{eff}}_{(0,0)}(\varphi=0)=-\frac14$ (leading to an effective central
charge $c^{\mathrm{eff}}=3$).
%
%----------------------------------------------------------------------
%
For odd $L$ the lowest state in the NS sector is doubly degenerate.  One of
the Bethe root configurations for this state consists of $N_\pm=(L\pm1)/2$
type-$(\pm)$ strange strings, i.e.\ $\Delta N=1$.  Its scaling dimension
exhibits a logarithmic dependence on the system size, $X^*=-\frac14+O((\Delta
N/\log L)^2)$ \cite{EsFS05,FrMa11}, consistent with the emergence of a
continuum of levels starting at $X^{\mathrm{eff},NS}(0,0) =-\frac14$ in the
thermodynamic limit.

\begin{figure}[t]
  \includegraphics[width=0.6\textwidth]{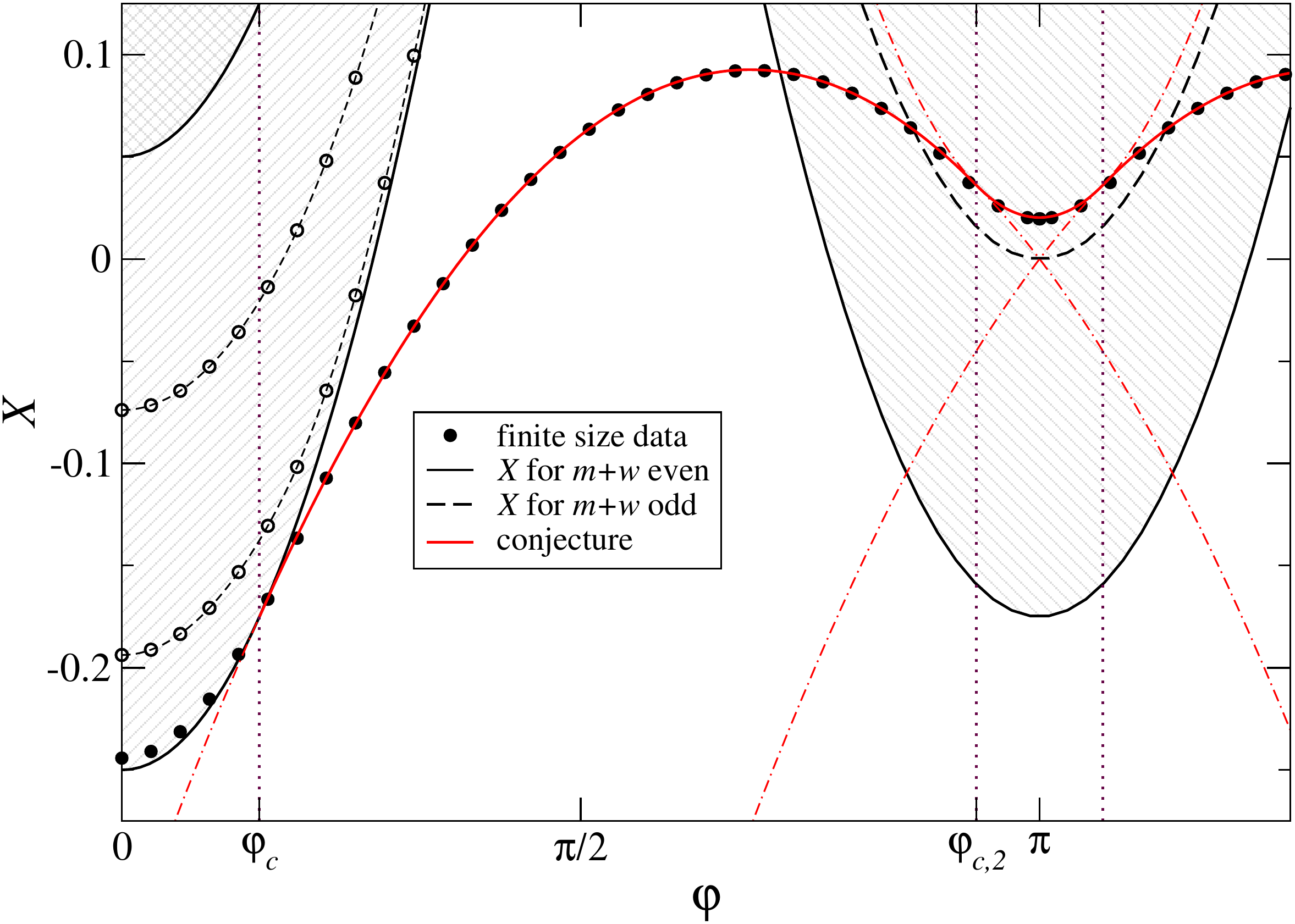}
  \caption{Evolution of the lowest states in the NS sector of the superspin
    chain of odd length $L$ with the twist angle $\alpha=-\pi+\varphi$:
    bullets are the scaling dimensions obtained from the solution of the Bethe
    equations (\ref{eq:BAE}) for $L=27$ evolving from a root configuration
    consisting of 
    strange strings with $|\Delta N|=1$ for $\varphi=0$, i.e. the NS ground
    state, for anisotropy $\gamma=17\pi/40$ .  Open circles show the flow of
    scaling dimensions with $|\Delta N|=3$ and $5$ for the same
    parameters. Black lines indicate the lowest effective scaling dimensions
    $X^{\mathrm{eff}}_{(m,w)}(\varphi)$ of the superspin chain, i.e.\
    $(m,w)=(0,0)$, $(1,-1)$, $(2,0)$, and $(0,-1)$, for this anisotropy.  The
    shaded areas indicate 
    the observed continua of scaling dimensions starting at
    $X^{\mathrm{eff}}_{(m,w)}(\varphi)$ with even $m+w$.  The conjectured
    $\varphi$-dependence of the discrete level $X^*_{(0,0)}$,
    Eq.~(\ref{discrete-conj}), is shown in red (dash-dotted lines indicate
    continuations of the functions appearing in the piecewise definition of
    $X^*_{(0,0)}(\varphi)$ beyond their domain of definition).
    \label{fig:specflow}}
\end{figure}

\begin{figure}[t]
  \includegraphics[width=0.6\textwidth]{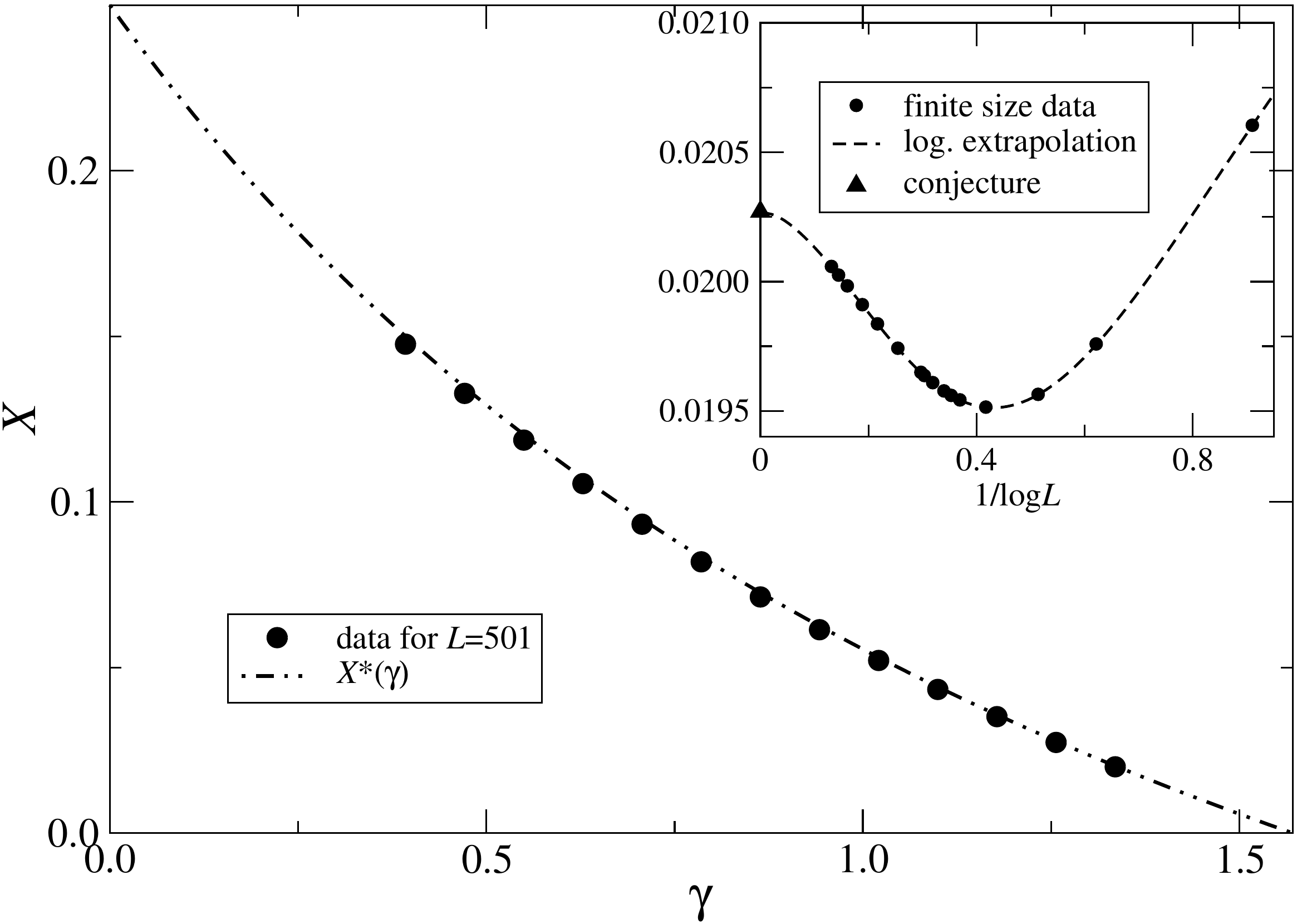}
  \caption{Scaling dimension of the discrete level in the Ramond
    sector, $\varphi=\pi$, as a function of the anisotropy parameter
    $\gamma$: bullets are the numerical data for system size $L=501$,
    dash-dotted line is the conjectured value (\ref{xstar-conj}) in
    the thermodynamic limit.  The inset shows the $L$ dependence of
    the finite size data ($\bullet$) for $\gamma=17\pi/40$ together
    with the extrapolation based on an assumed rational dependence of
    the data on $1/\log L$ (dashed line).
    \label{fig:xstar}}
\end{figure}

Adiabatically changing the twist angle we can follow this state under the
spectral flow.  For $|\varphi| < \varphi_c= \pi/k$ %\pi-2\gamma$
its scaling dimension $X^*_{(m,w)=(0,0)}(\varphi)$ stays within the continuum
above $X^{\mathrm{eff}}_{(0,0)}(\varphi)$.  As the twist approaches
$\pm\varphi_c$ the strange string with largest roots goes to $\infty$.  Beyond
$\pm\varphi_c$ the root configuration changes and the finite size scaling
dimension of the state deviates from (\ref{eq:fsspec}).  Unlike the higher
excitations ($|\Delta N|>1$) within this class of Bethe states it splits off
from the emerging continuum above $X^{\mathrm{eff}}_{(0,0)}(\varphi)$, see
Fig.~\ref{fig:specflow}.  Based on our finite size data we conjecture the
following $\varphi$-dependence of the scaling dimension of the scaling
dimension:
\begin{equation}
  \label{discrete-conj}
  \begin{aligned}
    X^*_{(0,0)}(\varphi) &= \begin{cases}
      -\frac14 + \frac{k}{2}\left(\frac{\varphi}{\pi}\right)^2
      - \frac{2k-1}{(k-1)^2}\,
         \left(\frac12 -
           \frac{k}{2}\left|\frac{\varphi}{\pi}\right|\right)^2\,
         & \mathrm{for~}
            \varphi_c\le|\varphi|\le \varphi_{c,2}
              %\pi-\frac{\pi(k-1)}{k(2k-1)}
            \,\\
      \frac{k}{2}\left(1-\frac{\varphi}{\pi}\right)^2
      + \frac14\,\frac{1}{2k-1}
         & \mathrm{for~} |\varphi-\pi| \le | \varphi_{c,2} -\pi |
            % \frac{\pi(k-1)}{k(2k-1)}
         \,
     \end{cases}
   \end{aligned}
\end{equation}
where $\varphi_{c,2} = \pi-\frac{\pi(k-1)}{k(2k-1)}$.
%----------------------------------------------------------------------
%----------------------------------------------------------------------
%
Hence we find that one state from the continuum above
$X^{\mathrm{eff},NS}(0,0)$ evolves, under the spectral flow
$\varphi=0\ldots\pi$, to a discrete level with dimension
\begin{equation}
  \label{xstar-conj}
  X^*_{(0,0)}(\pi) = \frac14\, 
    \frac{\pi-2\gamma}{\pi+2\gamma}\,
    = \frac14\,\frac{1}{2k-1}\,
\end{equation}
in the Ramond sector. The Bethe root configuration for this state for
$\varphi=\pi$ consists of $(L-1)/2$ of the usual $2$-strings (consisting of
two complex conjugate rapidities from the same level \cite{TaSu72}) and, in
addition, one single root at $\infty$ on either level, i.e.\footnote{This
  configuration has already been observed for $L=3$ in Ref.~\cite{FrMa11} but
  not considered further.}
\begin{equation}
  \lambda_j^{(1)}=-\lambda_j^{(2)} \in 
  \left\{\mu_k^\pm\,:\,
    \mathrm{Im}(\mu_k^\pm) \simeq\pm\frac{\gamma}{2}\,,
    \,k=1\ldots \frac{L-1}{2} \right\}
  \cup \left\{\infty\right\}\,.
\end{equation}
Note that there are strong logarithmic finite size corrections to
scaling to (\ref{xstar-conj}), similar as for the states in the
continuum,
see Figure~\ref{fig:xstar}.
Increasing the twist further we find that $X^*_{(0,0)}(\varphi) =
X^*_{(0,0)}(2\pi-\varphi)$.  For $\varphi>2\pi-\varphi_{c,2}$ this
level coincides with
\begin{equation}
  X^*_{(0,-2)} = X_{(0,-2)}^{\mathrm{eff}}(\varphi)
  - \frac{2k-1}{(k-1)^2} 
  \left(\frac12 - \frac{k}{2}\left|\frac{\varphi}{\pi}-2\right|\right)^2\,,
\end{equation}
and disappears in the continuum above 
$X^{\mathrm{eff}}_{(0,-2)}(\phi)$ at $\varphi=2\pi-\varphi_c$.

Concluding our analysis of the scaling dimensions under the spectral flow we
note that starting from the lowest state in the continuum above
$X^{\mathrm{eff}}_{(1,-1)}(\varphi=\pi)$ (for even $L$) we have observed
another discrete level in the spectrum of the superspin chain
\begin{equation}
  X^*_{(1,-1)} (\varphi) = X_{(1,-1)}^{\mathrm{eff}}(\varphi)
  - \frac{2k-1}{(k-1)^2} 
  \left(1 - \frac{k}{2}\left|\frac{\varphi}{\pi}-1\right|\right)^2\,,
  \quad |\varphi-\pi|>\frac{2\pi}{k}\,.
\end{equation}

%\section{Quasi momentum of the discrete state}
%
Let us now analyze the quasi momentum of the state with dimension $X^*$
(\ref{discrete-conj}) identified above.  An operator similar to
(\ref{eq:quasimom}) has first been introduced for a $Z_2$-staggered six-vertex
model where its (real) eigenvalues have been identified with the quantum
number parametrizing the spin $j$ of the $SL(2,\mathbb{R})$ affine primaries
from the continuous series, $j=-1/2 + iK$ \cite{IkJS12}.  Based on this
identification the staggered six-vertex model has be shown to be described by
the $SL(2,\mathbb{R})/U(1)$ Euclidean black hole CFT in the continuum limit.

\begin{figure}[t]
  \includegraphics[width=0.6\textwidth]{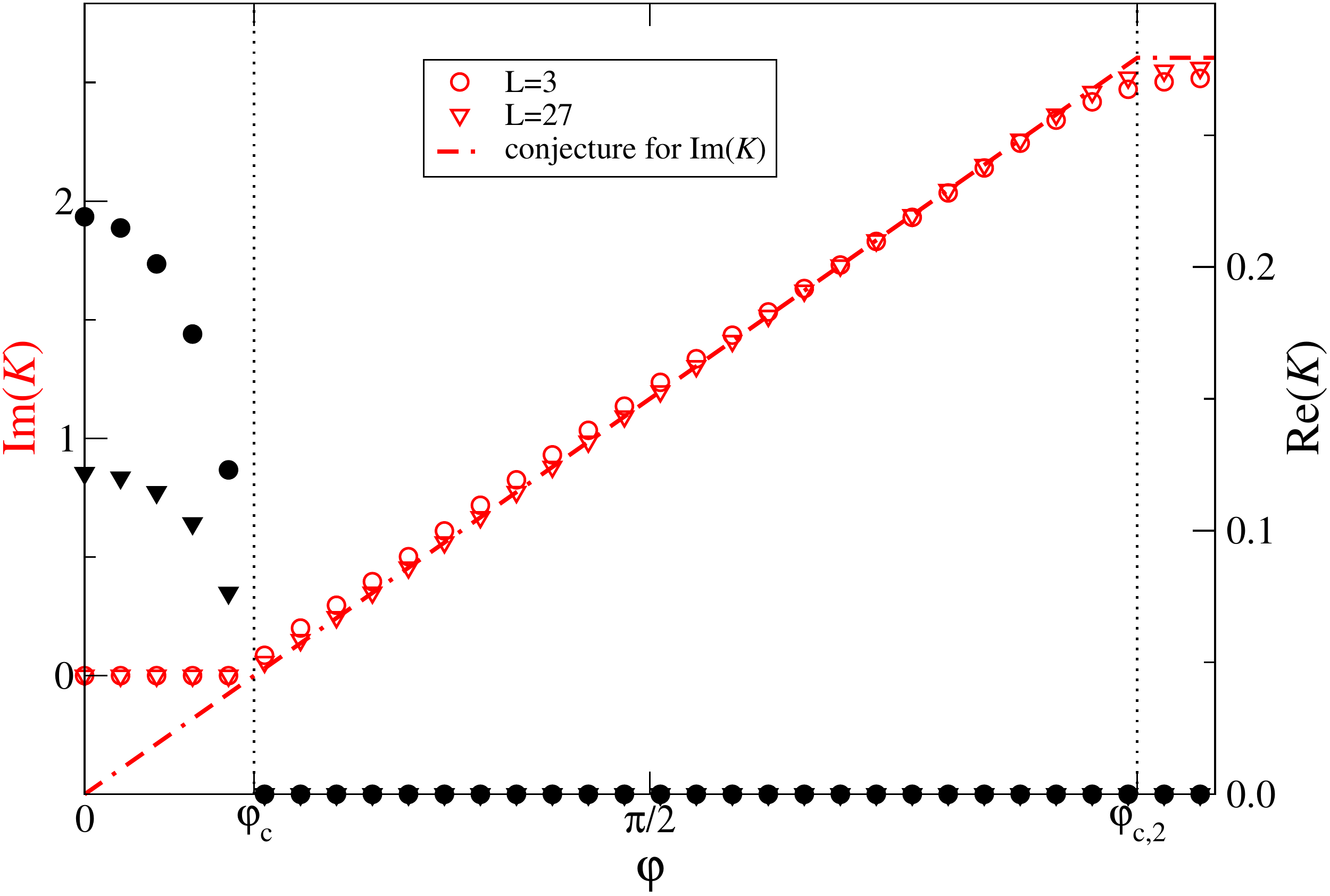}
  \caption{Quasi momentum of the discrete level for $\gamma=17\pi/40$ vs.\
    twist angle $\varphi$: filled (open) symbols are the numerical data for
    the real (imaginary) part of $K$ for system sizes $L=3$ and $27$.
    Dash-dotted line (in red) is the conjecture for $\mathrm{Im}(K)$,
    Eq.~(\ref{kstar-conj}), saturating at $\varphi=\varphi_{c,2}$.
    \label{fig:kstar}}
\end{figure}

For the superspin chains considered here the quasi momentum allows to label
the continua of scaling dimensions in a similar way \cite{FrHoXX}: in fact
starting with the lowest state in the NS sector but outside of the XXZ
subspace the corresponding eigenvalue $K$ is found to take real values for
$|\varphi|<\varphi_c$, see Figure~\ref{fig:kstar}.  
The amplitude vanishes as $1/\log L$ in the thermodynamic limit.  This agrees
with the density of states observed in the continua \cite{EsFS05,FrMa11}.

For $|\varphi|>\varphi_c$, however, i.e.\ as the discrete level
(\ref{discrete-conj}) leaves the continuum of scaling dimensions, the
quasi momentum becomes imaginary with a
linear dependence on the twist angle $\varphi$
\begin{equation}
  \label{kstar-conj}
  K^*(\varphi) = \frac{i}{2(\pi-2\gamma)}\,(|\varphi|-\varphi_c)
  = i\left(\frac{k|\varphi|}{2\pi}-\frac12\right) \,
  \quad \mathrm{for~} \varphi_c\le \varphi \le \varphi_{c,2}\,
\end{equation}
(we have chosen the branch of the logarithm in (\ref{eq:quasimom}) such that
$\mathrm{Im}(K) \in [0,(k-1)/2]$.)  Corrections to scaling in this expression
are small which allows to observe this behaviour already for $L=3$, see
Figure~\ref{fig:kstar}.
As for the scaling dimension (\ref{discrete-conj}) the $\varphi$-dependence of
the quasi momentum changes when the twist is increased beyond
$|\varphi|=\varphi_{c,2}$: while $K$ remains purely imaginary its value
saturates at
\begin{equation}
  \label{kstar-bound}
  \mathrm{Im}(K^*(\varphi)) = \frac{4\gamma^2}{\pi^2-4\gamma^2} =
  \frac{(k-1)^2}{2k-1} \, \quad
  \mathrm{for~} \varphi\ge \varphi_{c,2}\,.
\end{equation}
Note that $\mathrm{Im}(K^*(\pi))$ is smaller than the maximum
$\gamma/(\pi-2\gamma) = (k-1)/2$ which is possible according to the definition
(\ref{eq:quasimom}).

The emergence of a discrete level out of the continuum at finite twist is
strongly reminiscent to the situation in the $a_2^{(2)}$ spin chain
\cite{VeJS14}: in the black hole CFT describing the low energy behaviour of
this lattice model it is understood as a consequence of the inclusion of
principal discrete representations of $SL(2,\mathbb{R})$ as Kac-Moody
primaries in the coset theory.  For operators corresponding to normalizable
states in the parent WZNW theory of the model the spin of these representations
(related to the momentum along the non-compact direction of the target space,
an infinite cigar for the $SL(2,\mathbb{R})/U(1)$ sigma model) is restricted
to values $j\le-1/2$.  To satisfy this bound a finite, non-zero twist has to
be applied \cite{HaPT02,RiSc04}.
For the bosonic $SL(2,\mathbb{R})/U(1)$ coset at level $k$ the spectrum of
discrete representations also needs to be truncated by the 'unitarity bound'
$j\ge-(k-1)/2$ to guarantee non-negative conformal weights
\cite{DiVV92,MaOo01,HaPT02}.

The appearance of the discrete level $X^*$ (\ref{discrete-conj}) in the
spectrum of the superspin chain at twist $\varphi_c$ accompanied with the
change of quasi momentum from real to imaginary can be interpreted in a
similar way: states in the continuum of scaling dimensions have real $K$, the
discrete levels are characterized by imaginary quasi momentum.  The observed
bound $\mathrm{Im}(K)\ge 0$ for states in the lattice model can be attributed
to the normalizability of the corresponding primaries.  Continuing
(\ref{kstar-conj}) to $|\varphi|<\varphi_c$ would yield imaginary quasi
momenta $-1/2 \le \mathrm{Im}(K) < 0$ which is not realized in the spectrum of
the superspin chain.
As for a restriction of $\mathrm{Im}(K)$ from above (similar to the unitarity
bound in the string theory) our data for the lattice model do not provide a
conclusive answer.  We would need to see a level being 'absorbed' by the
continuum at such a bound under the spectral flow (as is the case for the
$a_2^{(2)}$ spin chain \cite{VeJS14}).  It would be tempting to associate this
bound with $\mathrm{Im}(K) \le (k-1)/2$ as implied by our choice of the branch
for the logarithm in (\ref{eq:quasimom}).  Since the quasi momentum for the
discrete level studied above saturates at the value (\ref{kstar-bound}) below
$(k-1)/2$, however, our data cannot support this conjecture.

%######################################################################
%######################################################################
%######################################################################
%######################################################################

%\section{Conclusions}\label{sec:conclusiuons}
To summarize, we have studied the spectral flow for a staggered superspin
chain with continuous and discrete components in the spectrum of scaling
dimensions under a twist in the boundary conditions.  Based on the exact
solution of this model we identified a state from the continuous part of the
spectrum in the Neveu-Schwarz sector (i.e.\ with anti-periodic boundary
conditions for the fermionic degrees of freedom) which under variation of the
twist flows to a discrete level in the Ramond sector which has not been
identified previously.  Levels can be attributed to the continuous or discrete
part of the spectrum based on their quasi momentum.  The appearance of
discrete levels in the spectrum of the lattice model has been argued to be
related to the normalizability of primary fields in the non-rational conformal
field theory describing the continuum limit.

\begin{acknowledgments}
  This work has been carried out within the research unit
  \emph{Correlations in Integrable Quantum Many-Body Systems}
  (FOR2316).  Funding by the Deutsche Forschungsgemeinschaft under
  grant No.\ Fr~737/9-1 is gratefully acknowledged.
\end{acknowledgments}

%\bibliography{base,books,frahm}

%
\end{document}